\documentstyle[twoside,fleqn,espcrc2,epsfig,amssymb]{article}

\def\vev#1{\langle #1 \rangle}

\newcommand{\bea}{\begin{eqnarray*}}
\newcommand{\eea}{\end{eqnarray*}}

\newcommand{\cO}  {{\mathcal O}}

\title{Super-heavy Quarkonia and the Gluon Condensate}

\author{Jochen Fingberg\\
        University of Wuppertal
        and CCRL,~NEC~Europe~Ltd.,
        53757~St.Augustin,~Rathausallee~10,~Germany}

\begin{document}

\begin{abstract}
The early idea that a non-perturbative gluon condensate affects the
spectrum of heavy quarks is revisited in the light of modern
simulation techniques.
We evaluate the low lying spectrum of bound states of two heavy quarks
for large hypothetical quark mass, $m_Q > m_b$, using
non-relativistic QCD and compare with other models to
test the consistency.
\end{abstract}

\maketitle

\noindent
The concept of a non-perturbative gluon condensate was initially introduced
by Shifman, Vainstein, and Zakharov \cite{SVZ}. A non-vanishing gluon
condensate can give rise to a shift of energy levels of heavy quarkonia
by second order Stark splitting.
Leutwyler and Voloshin \cite{VoloshinLeutwyler}
calculated the level shifts in leading order perturbation theory.
Recent improvements like the stochastic vacuum model SVM \cite{DoSi89},
non-perturbative field strength correlators \cite{DiGiacomo} or a
combination of the SVM with lattice calculations \cite{Bali97,Brambilla97})
try to solve some of the problems of the original LV-model.
Using NRQCD has the advantage that no apriori knowledge about the 
heavy quark potential or the vacuum background field is needed.
The physical problem is reduced to a simulation of non-relativistic
heavy quarks in a correlated, relativistic gluonic background.
We use an $\cO (a^4)$ improved gauge action with a plaquette and a
$1 \times 2$ Wilson loop term for the relativistic simulations and
an $\cO (v^4,~v^6)$ corrected non-relativistic action
\[
S_{NR}=\psi^\dagger~\left(D_t + H_0 + \delta H(c_i)\right)~\psi
\]
with tree level couplings, $c_i=1$, for the heavy quarks.
The quark's Green's functions are calculated from the standard
(asymmetric in $\delta H$) evolution equation.
Meson propagators with specific quantum numbers are obtained
from the quark propagators using a smearing procedure with
trial wave-functions from a solution of the radial Schr\"odinger equation
using a method from ref.~\cite{Armin}
for each value of the bare quark mass. Preliminary results based on Coulomb
wave-functions and a small subset of the configurations
were already reported at a previous conference \cite{lat96}.

For the simulations we use a
lattice of size $16^4$ and a gauge coupling of $\beta=9.17$
to obtain a very fine spacing necessary to resolve small bound states.
After generating 360 decorrelated configurations on a Quadrics Q4
we fix to Coulomb gauge with a stopping criterion,
$\theta < 10^{-6}$. The simulations require
60 days of Quadrics Q4 time,
70 days on a workstation for gauge-fixing, and
more than 150 days of CM2/CM5 time.
We apply standard correlated multi-state multi-exponential fits
with vector fits for a set of S states
and ratio fits for hyperfine and S-P splittings.
The error is determined by a jackknife procedure including
the uncertainty from the fit range, $[t_{min},~N_{\tau}]$.
A consistent scale is obtained from $V_{q\bar q}$
and the force indicating that scaling violations are
indeed small at this high value of the gauge coupling
(see ref.~\cite{JF} for details).

As a first step our results can be compared to
leading order perturbation theory which allows to separate the
Coulomb from the condensate contribution. Following refs.~\cite{TY,STY}
the level splittings are given to lowest order by the expressions
\bea
\Delta E_{hfs}\!&\simeq&\!\frac{m(C_F\alpha_s)^4}{3}
\!+\!\frac{.5425\pi C_F^4}{m^3\alpha_s^2}~f_{10}~\vev{\alpha_sG^2}\\
\Delta E_{SP}\!&\simeq&\!\frac{m(C_F\alpha_s)^2}{16/3}
\!+\!\frac{64\pi}{C_F^4m^3\alpha_s^4}~f_{21}~\vev{\alpha_s G^2}\\
f_{nl}^{-1}\!&=&\!1+(\lambda_G^{-1}/[m(4/3\alpha_s)^2])~\rho_{nl}\\
\rho_{10}\!&=&\!2.48,~~~~\rho_{21}=11.20,~~~~C_F=4/3~~~~.
\eea
There are two length scales in the system: the gluon correlation
length $\lambda_G$ and the quark correlation length $\lambda_Q$
which is related to the rotational period of the two heavy quarks
inside the quarkonium, $\lambda_Q \simeq 2\pi\sqrt{\vev{r^2}/\vev{v^2}}$.
The shortest correlation length governs the 
behaviour of the energy levels. There are two limiting cases:
a static condensate ($\lambda_G >> \lambda_Q$) when $f_{nl}=1$ and a
rapidly varying condensate ($\lambda_G << \lambda_Q$) when $f_{nl}<1$.

We can solve the set of equations \{$\Delta E_{hfs}$, $\Delta E_{SP}$\}
for the two unknowns $\alpha_s$ and $\vev{\alpha_s G^2}$
inserting the values for the splittings from NRQCD shown in
figs.~1 and 2.
\begin{figure}[t]
\epsfig{file=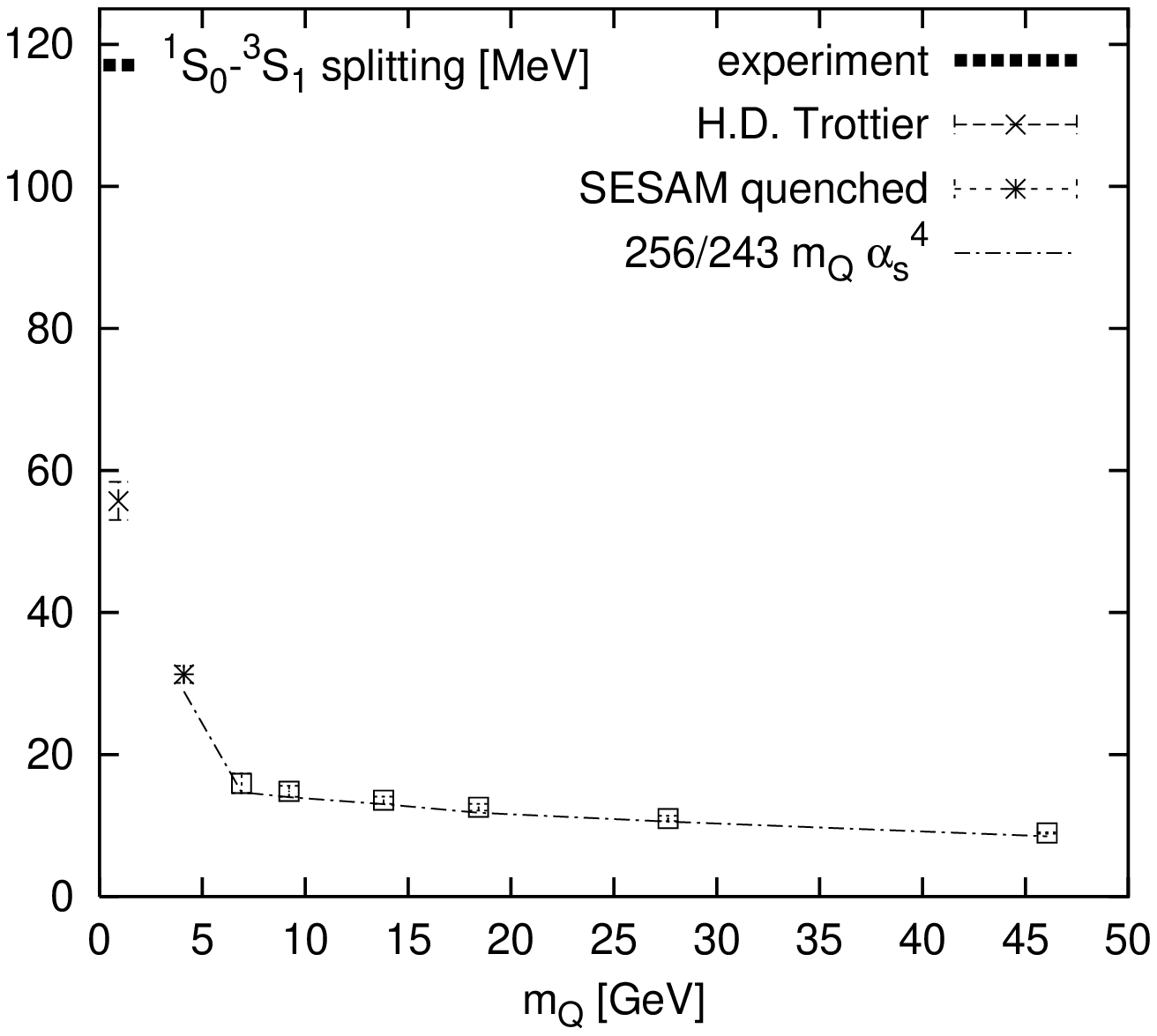,bbllx=75,bblly=62,bburx=453,bbury=404,width=\linewidth}
{\vskip 0.4\baselineskip\parbox[c]{\linewidth}
  {Figure 1.~The hyperfine splitting between the ground state of
   the ${}^3S_1$ and the ${}^1S_0$ meson. The line gives the perturbative
   Coulomb contribution with $\alpha_s(m_Q)$ as shown in fig.~3.
   Also shown are the $c\bar c$ \cite{trottier} and $b\bar b$
   values \cite{sesam}.}
}
\end{figure}
\begin{figure}[t]
\epsfig{file=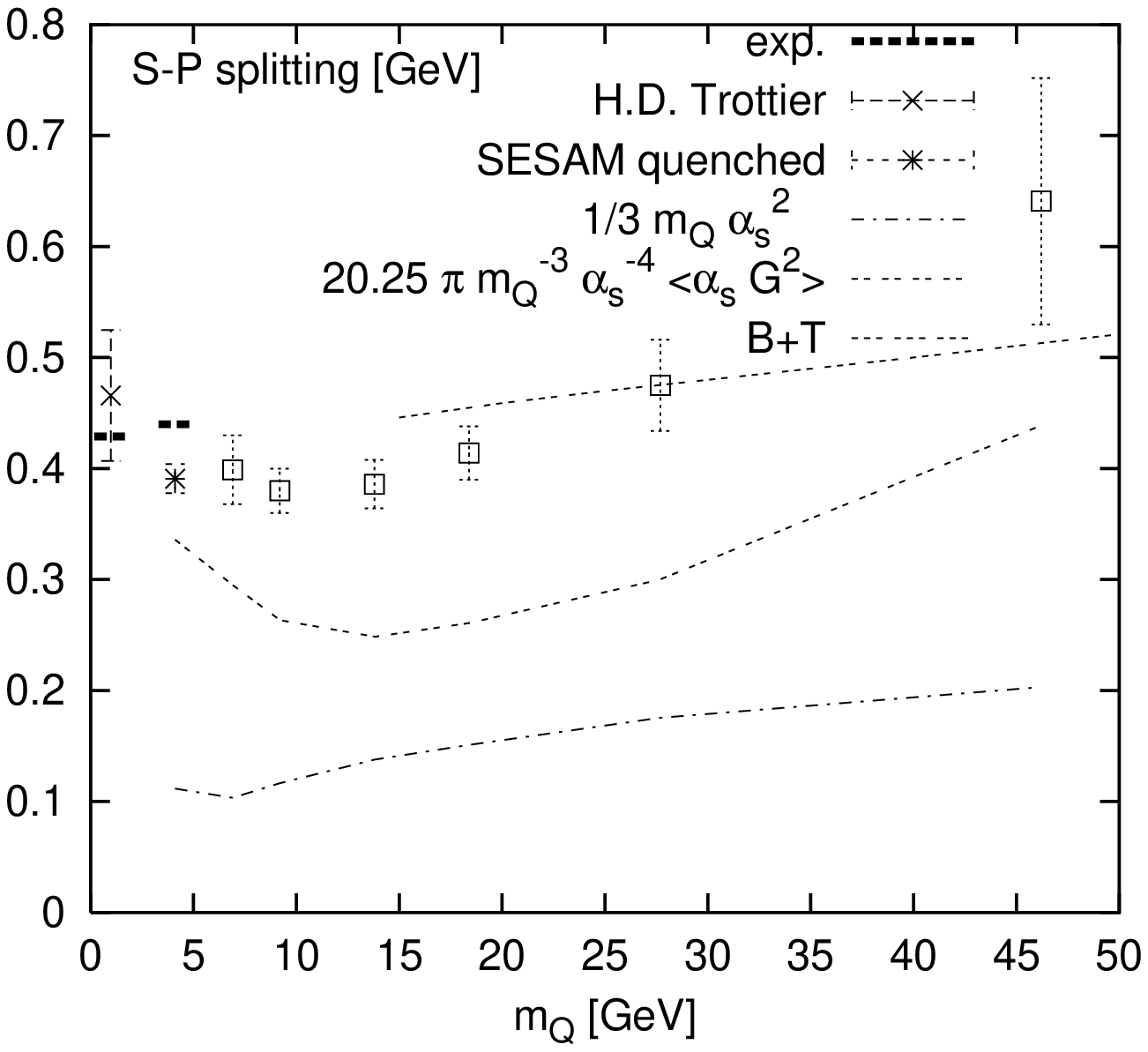,bbllx=75,bblly=62,bburx=453,bbury=404,width=\linewidth}
{\vskip 0.4\baselineskip\parbox[c]{\linewidth}
  {Figure 2.~The S-P splitting from NRQCD (squares),
   perturbation theory and for the Buchm\"uller and
   Tye model \cite{BMT}.}
}
\end{figure}
In this way we obtain the strong coupling and the condensate as a function
of the bare quark mass shown in figs.~3 and 4.
\begin{figure}[htb]
\epsfig{file=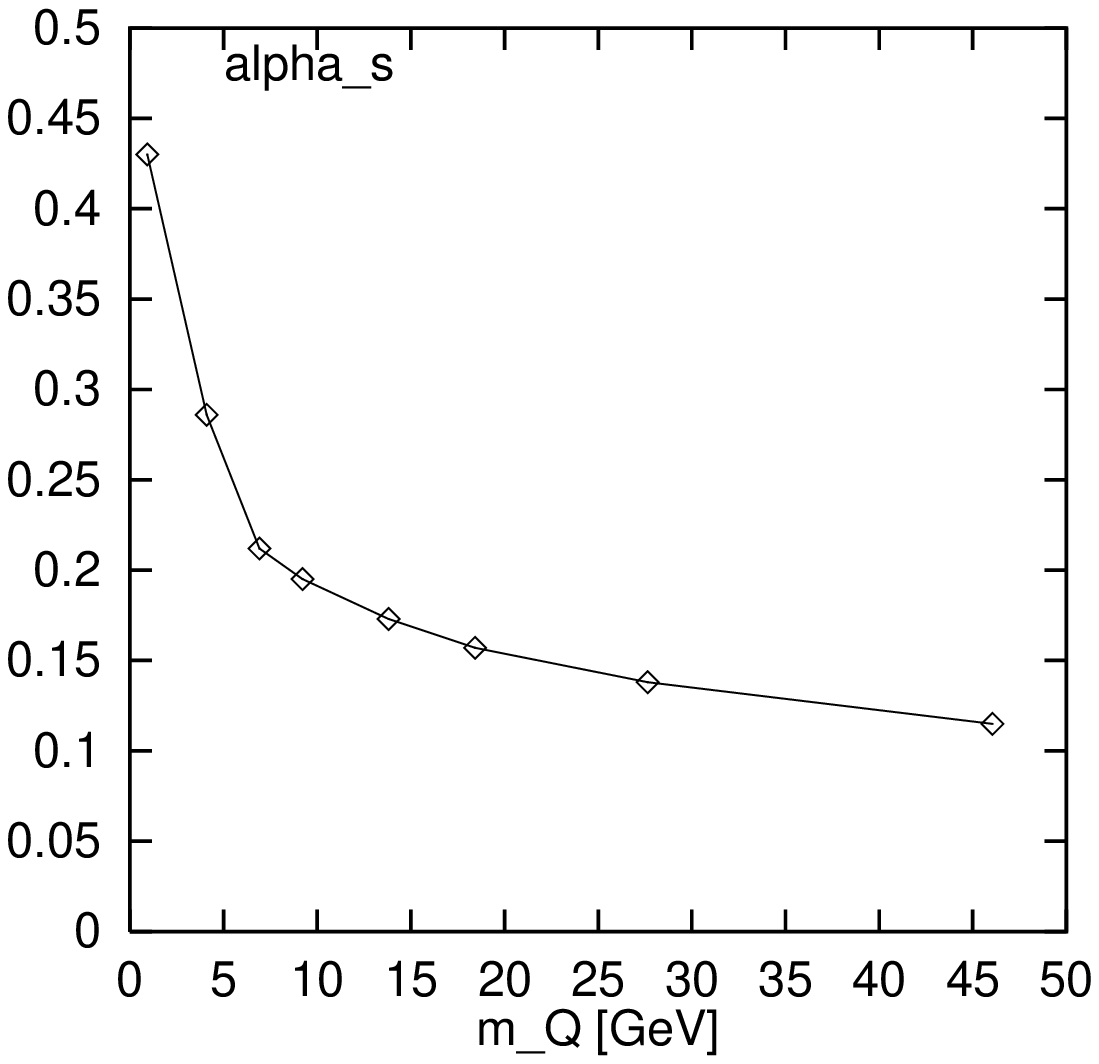,bbllx=80,bblly=48,bburx=396,bbury=350,width=\linewidth}
{\vskip 0.4\baselineskip\parbox[c]{\linewidth}
  {Figure 3.~The running coupling $\alpha_s$
   from a comparison of the S-P and
   the hyperfine splitting of the ground state from NRQCD
   with perturbative predictions.}
}
\end{figure}
\begin{figure}[htb]
\epsfig{file=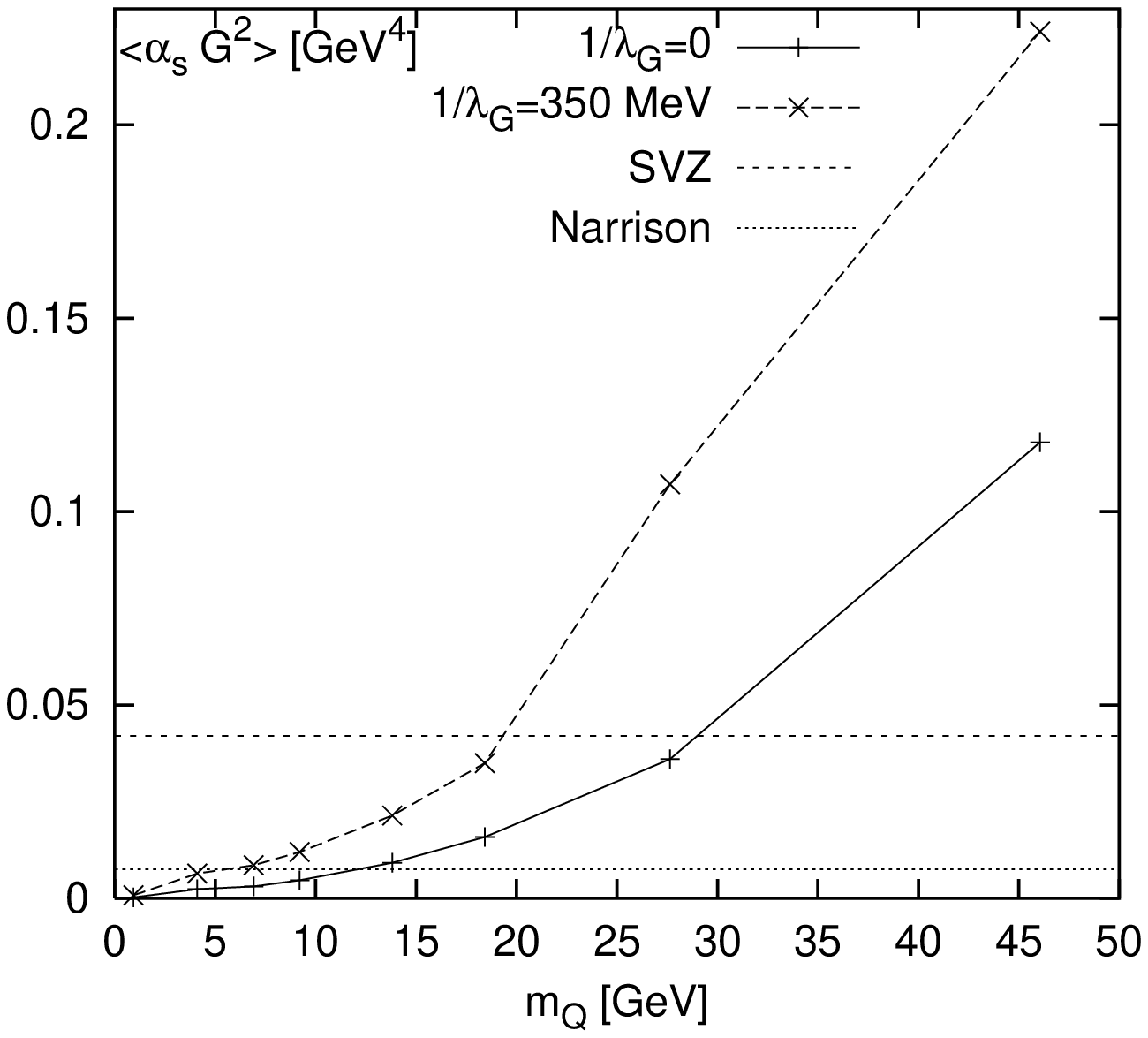,bbllx=80,bblly=62,bburx=452,bbury=396,width=\linewidth}
{\vskip 0.4\baselineskip\parbox[c]{\linewidth}
  {Figure 4.~A comparison of the gluon condensate
   for the static case, $\lambda_G \rightarrow \infty$ and for
   $\lambda_G^{-1}=350$ MeV with different values from the
   literature (SVZ: \cite{SVZ}, Narrison: \cite{Narrison}).}
}
\end{figure}
If we reverse this procedure for a while and
accept that fig.~3 describes a reasonable behaviour of the running
coupling $\alpha_s(m_Q)$ then the perturbative Coulomb contribution
($\simeq 1.05~m_Q \alpha_s^4$) shown in fig.~1
as a line together with our data
completely dominates the hyperfine splitting $E_{hfs}$. On the other hand
for the S-P splitting the condensate contribution is larger than the
Coulombic part which has been criticised before and sheds doubt on the
treatment of the condensate as a perturbation. The magnitude of the
contributions to
$E_{hfs}$ and $E_{SP}$ does not change significantly
as the gluon condensation
length is varied. The parameter $\lambda_G$ merely rescales the value of
the condensate by a factor $f_{nl}(\lambda_G)$. If we nevertheless
take the perturbative results seriously we see from fig.~4 that it is possible
to get a value of the condensate compatible with sum rule results.
However, the value of  $\vev{\alpha_s G^2}$ strongly varies with quark mass.
The static approximation ($\lambda_G \rightarrow \infty$) should be
valid for large quark mass while the approximatiuon of a rapidly varying
condensate ($\lambda_G << \lambda_Q$) is expected to hold for intermediate
masses, $m_Q<40$~GeV, because $\lambda_Q$ is decreasing with $m_Q$.

\section*{Summary}
At large $m_Q$ the hyperfine splitting of the ground state
becomes compatible with Coulombic behaviour. However, the
S-P splitting from NRQCD is neither in accord with the B+T model nor
with leading order perturbative results for LV-type models.
On the level of lowest order perturbation theory
no consistent picture of the effect of the gluon condensate
can be found allthough there is evidence that the general trend
predicted by the analytic results is qualitatively correct.
The NRQCD hyperfine splitting of the ground state is decreasing and the S-P
splitting is increasing with $m_Q$ for large quark mass consistent
with  the perturbative dependence on different powers of $\alpha_s$.
It will be interesting to see \cite{JF} if a comparison with more rigorous 
models like the combined SVM and lattice approach will consolidate
the situation. An alternative might be a closed lattice approach.
Then one is faced with the difficult problem to switch the condensate
on and off during the simulations.

\section*{Acknowledgement}
J.F. was supported by a DFG fellowship during the initial phase of the 
project. We also wish to thank the IAI
at the University of Wuppertal and the computer centre of the University
of Erlangen for their support.


\begin{thebibliography}{99}
\bibitem{SVZ}
V.I.~Zakharov, (1978);
M.A.~Shifman, A.I.~Vainstein and V.I.~Zakharov,
Nucl. Phys. B147 (1979) 385 and 448.

\bibitem{VoloshinLeutwyler}
M.B.~Voloshin, Nucl. Phys. B154 (1979) 365;
H.~Leutwyler, Phys. Lett. B98 (1981) 447.

\bibitem{DiGiacomo}
A.~Di~Giacomo and H.~Panagopoulos, Phys. Lett. B285 (1992) 133.
 

\bibitem{BMT}
W.~Buchm\"uller and S.-H.H.~Tye, Phys. Rev. D24 (1981) 132;
J.L.~Rosner et al., Phys. Rev. D53 (1996) 2742.


\bibitem{DoSi89}
H.G.~Dosch and Yu.A.~Simonov,
Phys. Lett. B209 (1988) 339;
Z. Phys. C45 (1989) 147.


\bibitem{Bali97}
G.~Bali, K.~Schilling and A.~Wachter,
Phys. Rev. D56 (1997) 2566.

\bibitem{Brambilla97}
G.~Bali, N.~Brambilla and A.~Vairo,
eprint-archive hep-lat/9709079, December 1997.


\bibitem{TY}
S.~Titard and F.J.~Yndurain,
Phys. Rev. D49 (1994) 6007;
Phys. Lett. B351 (1995) 541;
Phys. Rev. D51 (1995) 6348.

\bibitem{STY}
Yu.A.~Simonov, S.~Titard and F.J.~Yndurain, Phys. Lett. B354 (1995) 435.

\bibitem{lat96}
J.~Fingberg, Nucl. Phys. Proc. Suppl. 53 (1996) 405.

\bibitem{trottier}
H.D.~Trottier, Phys. Rev. D55 (1997) 6844.

\bibitem{sesam}
N.~Eicker, T.~Lippert, K.~Schilling, A.~Spitz , J.~Fingberg,
S.~Gusken, H.~Hoeber, J.~Viehoff, HLRZ-1997-35, Sep 1997.

\bibitem{Armin}
A.~Wachter, PhD-thesis, Universit\"at Wuppertal, WUB-DIS 96-21,
November 1996.

\bibitem{Narrison}
S.~Narrison, Phys. Lett. B 1996.

\bibitem{JF}
J.~Fingberg, work in progress.
\end{thebibliography}
\end{document}